\documentclass[%
 reprint,
 amsmath,amssymb,
 aps,
]{revtex4-2}

\usepackage{graphicx}
\usepackage{dcolumn}
\usepackage{bm}
\usepackage{subcaption}
\usepackage{hyperref}
\usepackage{verbatim}   
\usepackage{physics}    
\usepackage{cleveref}   
\usepackage{xcolor}
\usepackage{hyperref}   

\begin{document}

\preprint{APS/123-QED}

\title{Including sample shape in micromagnetics with 3D periodic boundary conditions}

\author{Frederik Laust Durhuus}
\affiliation{
 Department of Energy Conversion and Storage, Technical University of Denmark - DTU, DK-2800 Kgs. Lyngby, Denmark
}%

\author{Andrea Roberto Insinga}
\affiliation{
 Department of Energy Conversion and Storage, Technical University of Denmark - DTU, DK-2800 Kgs. Lyngby, Denmark
}%


\author{Rasmus Bjørk}
 \email{rabj@dtu.dk}
\affiliation{
 Department of Energy Conversion and Storage, Technical University of Denmark - DTU, DK-2800 Kgs. Lyngby, Denmark
}%

\date{\today}

\begin{abstract}
Periodic boundary conditions (PBCs) for computing magnetic fields in repeating magnetic structures, e.g. in micromagnetic simulations, are typically imposed using the quasi periodic macrogeometry approach, where many copies of the simulated domain are introduced. This can be computationally problematic, especially if the simulated domain is incommensurate with the desired sample shape. In this work, we present a formal proof that for sufficiently large magnetic samples, only the average magnetisation gives non-negligible shape effects. Using this insight, we develop a simple, computationally efficient modification of existing implementations which incorporates shape effects in PBC methods. 
\end{abstract}

\maketitle

\section{Introduction}

Micromagnetism is a common approach to calculate the structure and dynamics inside magnetic materials. Micromagnetic simulations describe the simulated domain (here denoted $\Gamma$) as a collection of uniformly magnetised, interacting cells whose magnetic moments evolve in time by the Landau–Lifshitz–Gilbert (LLG) equation \cite{Donahue_1999,Vansteenkiste_2014,bjork_magtense_2021}. This relies on the fact that on a small enough length scale, exchange interactions ensure locally uniform magnetisation. The material specific single-cell length scale is typically on the order of nanometers, which means that micromagnetic simulations are generally restricted to microscopic domains. One way to bridge the gap between microscopic computations and macroscopic samples is by periodic boundary conditions (PBCs). Here one imagines the simulated domain surrounded by displaced copies with the exact same magnetisation distribution. Rather than treating the copies self-consistently, one only accounts for the field they produce within $\Gamma$ and exchange interactions at the surface of $\Gamma$. This greatly simplifies the calculation at the cost of assuming the simulated domain is fully representative of the entire magnet. 

In the popular OOMMF micromagnetics solver\cite{donahue_oommf_1999}, PBCs are implemented by having the domain copies repeat to infinity, but treating the more distant copies approximately. Above a critical distance the magnetic moments from the copies of each cell are modelled as smeared out into uniformly magnetised continua, as described in Refs.\ \cite{lebecki_periodic_2008,wang_two-dimensional_2010}.
The resulting integrals have been solved analytically in 1D (one-dimension)\cite{lebecki_periodic_2008} and 2D\cite{wang_two-dimensional_2010}, though they are not absolutely convergent in 3D. Other groups have implemented PBCs in 3D by Ewald summation\cite{wysocki_micromagnetic_2017}, a Fourier space method\cite{bruckner_strayfield_2021} or using Greens functions\cite{ai_periodic_2025}.

In 3D one encounters a fundamental issue however: the demagnetisation field depends on the shape of the whole magnet. For example, as in Ref.\ \cite{lebecki_periodic_2008}, consider a sphere and an ellipsoid which are uniformly magnetised and identical in all but shape. The demagnetisation field is $\vb{H}_\text{demag} = - \mathrm{N} \vb{M}$ where $\vb{M}$ is magnetisation and $\mathrm{N}$ is the shape-specific demagnetisation tensor. Because $\mathrm{N}$ is independent of volume, it does not matter how far the surface is from the simulated domain; $\vb{H}_\text{demag}$ will be different inside the two magnets. When applying PBCs in 3D out to infinity as in refs. \cite{wysocki_micromagnetic_2017,bruckner_strayfield_2021,ducevic_micromagnetic_2025}, one implicitly assumes that the system and by extension $\mathrm{N}$ are isotropic. Bruckner et al. highlight the exclusion of shape anisotropy as a strong point of their method\cite{bruckner_strayfield_2021}, and for the static susceptibility tensor $\chi$ of a multi-domain magnet or magnetic composite\cite{zambach_effect_2026}, the shape corrected value is simply $\chi_\text{eff} = [\mathrm{N} + \chi^{-1}]^{-1}$; at least when the demagnetisation field is approximately uniform.  

However, when one is interested in how sample shape affects dynamical properties, or in non-linear magnetic response such as most hysteresis curves, the shape dependent demagnetisation must be incorporated in simulations. The most standard solution is the macrogeometry approach\cite{fangohr_new_2009}, implemented e.g.\ in MuMax3\cite{vansteenkiste_design_2014}. Here, instead of true PBCs, there are a finite number of domain copies, forming a quasiperiodic system. Matsuo and Yamazaki demonstrated numerically that if the macrogeometry is large enough, the effect of distant domain copies is well-described by their average magnetisation $\vb{M}_\text{avg}$\cite{matsuo_demagnetizing_2011}. In fact, there are cases where the static susceptibility is well described by just $\vb{M}_\text{avg}$ and sample shape via $\mathrm{N}$ \cite{matsuo_demagnetizing_2011}. 

In this work, we prove formally that only the contribution from $\vb{M}_\text{avg}$ is scale-invariant, while the rest of the $H$-field does converge to a shape-independent value even in 3D, in agreement with Ref.\ \cite{matsuo_demagnetizing_2011}. Based on this realisation, we propose a simple method to account for sample shape in micromagnetism, which can both be used to reintroduce shape effects in simulations with PBCs to infinity and to modify quasi-periodic solutions yielding more flexibility in the choice of macrogeometry. Using the MagTense software for magnetostostatic and micromagnetic computations\cite{bjork_magtense_2021}, we test the method on a soft magnetic composite in a high-frequency, oscillating magnetic field. While this has previously been modelled e.g.\ in Refs.\ \cite{Wier_susceptibility_2022,Arzbacher_tomography_2015,bruckner_strayfield_2021,ducevic_micromagnetic_2025}, we present the first micromagnetic simulations of sample shape effects in the high-frequency regime.

\section{Theory}

\begin{figure}[htb]
    \centering
    \includegraphics[width=0.5\textwidth]{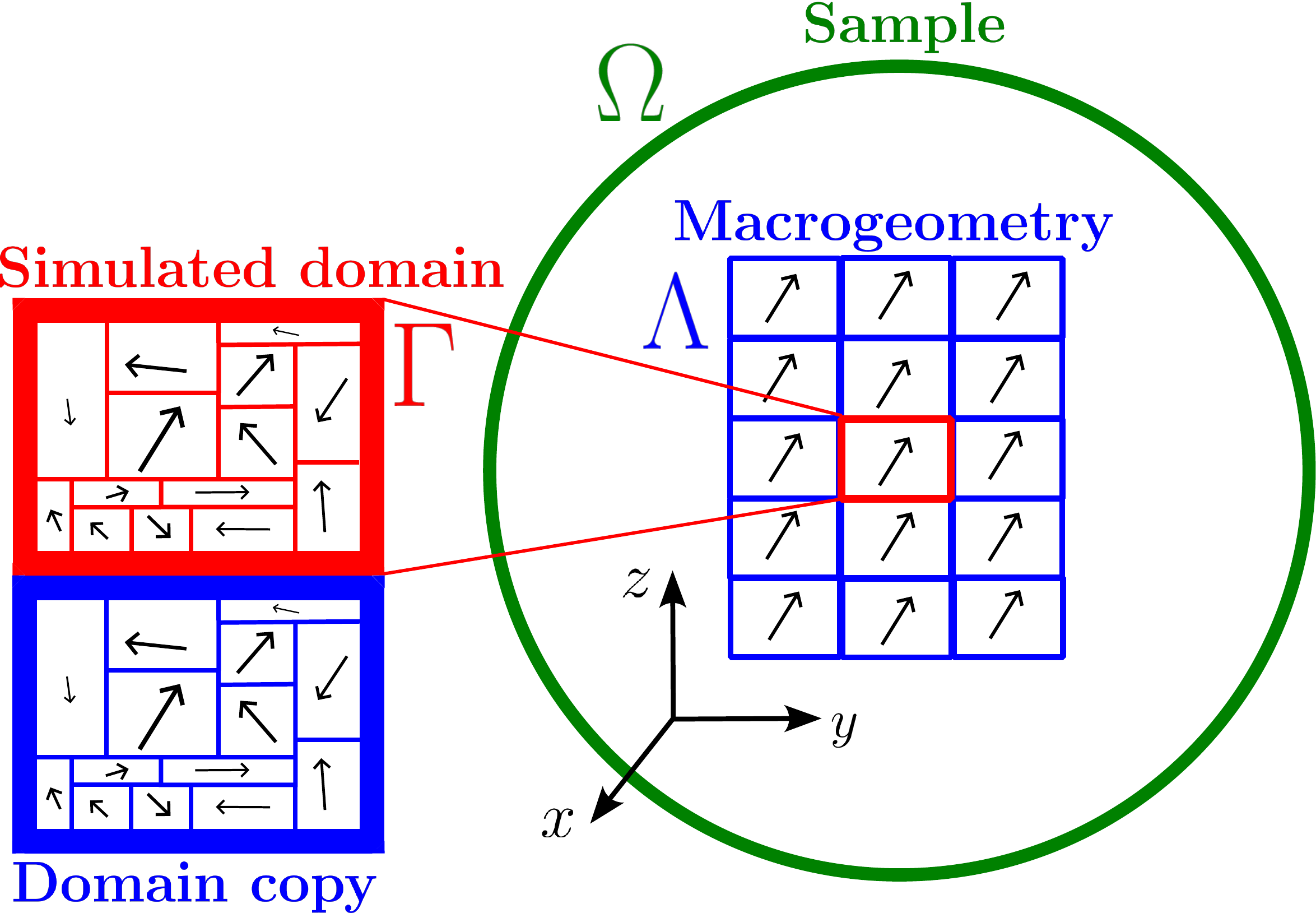}
    \caption{Illustration of micromagnetic problem in 3D with periodic boundary conditions. The interior of the green circle, $\Omega$, is the entire magnetic sample, the red region, $\Gamma$, is the simulated domain, and the blue regions are exact copies denoted $\Gamma_i$. The union of all the $\Gamma$-regions (red and blue rectangles) is denoted $\Lambda$. 
    For the purposes of the derivations presented here, the shape of $\Gamma$ and $\Omega$ as well as the number of $\Gamma$-copies in each direction is arbitrary.}
    \label{fig:PBC_3D_problem}
\end{figure}

\subsection{Implementation of PBCs in 3D}

We imagine $n_\text{copy}$ displaced, non-overlapping copies of the simulated domain, denoted $\Gamma_i$, which together form the region $\Lambda$ centered on $\Gamma$, as illustrated in \cref{fig:PBC_3D_problem}. Furthermore, let $c$ index the $n_\text{cell}$ cells within $\Gamma$ with magnetisations $\vb{M}_c$ and positions $\vb{r}_c$ relative to the center of $\Gamma$. We emphasize that $\Gamma$ can include zero-magnetisation cells, so besides compact magnets, the formalism also applies to arrays of magnetic particles or composites of magnetic and non-magnetic materials. In fact, even for disordered systems, if $\Gamma$ is large enough to include a representative distribution of magnetised grains and particles, PBCs may still be applicable.

The total magnetostatic field at position $\vb{r}$ from all the magnetisation within $\Lambda$ is given by
\begin{align}
    \vb{H}_\Lambda(\vb{r}) &= - \sum_{c} \mathrm{N}^\text{PBC}_{c}(\vb{r} - \vb{r}_c) \vb{M}_c,
\end{align}
with
\begin{align}
    \mathrm{N}^\text{PBC}_c(\vb{r} - \vb{r}_c) = \sum_i \mathrm{N}_{c}(\vb{r} - \vb{r}_c - \vb{r}_i),
\end{align}
where $\mathrm{N}_c$ is the demagnetisation tensor associated with cell $c$ and $\vb{r}_i$ is the center position of $\Gamma_i$. Thus, if the demagnetisation tensors are known for each cell in $\Gamma$, 
the field from the domain copies can be included by a simple sum performed once at the beginning of the simulation. 

Now, let $\Omega$ denote the entire sample volume and $\overline{\Omega} = \Omega - \Lambda$ be the difference of the two spaces. As shown in \cref{appsec:decomposing_magnetic_field} the magnetic field from this region can be written 
\begin{align}
    \vb{H}_{\overline{\Omega}} = \vb{H}_{\overline{\Omega}}^\text{avg} + \Delta \vb{H}_{\overline{\Omega}},
\end{align}
where $\vb{H}_{\overline{\Omega}}^\text{avg}$ is the field produced by $\overline{\Omega}$ if the region was uniformly magnetised with the same average magnetisation as $\Gamma$, i.e.\
\begin{align}
    \vb{H}_{\overline{\Omega}}^\text{avg}(\vb{r}) = \frac{1}{4\pi} \int_{\overline{\Omega}} \frac{3 [\vb{M}_\text{avg} \vdot (\vb{r} - \vb{r}')](\vb{r} - \vb{r}') - \vb{M}_\text{avg}}{\abs{\vb{r} - \vb{r}'}^3} \dd \vb{r}',
\end{align}
while $\Delta \vb{H}_{\overline{\Omega}}$ is the field contribution from locally non-uniform magnetisation. In \cref{appsec:3D_PBC_proof} we show that $\Delta \vb{H}_{\overline{\Omega}} \xrightarrow{} 0$ as $\Lambda$ grows to infinity. Hence, if the sample is large enough for $\Delta \vb{H}_{\overline{\Omega}}$ to converge before reaching the surface, the magnetostatic field inside $\Gamma$ is to a good approximation
\begin{align}
    \vb{H}(\vb{r}) &= \vb{H}_\Lambda(\vb{r}) + \vb{H}_{\overline{\Omega}}^\text{avg}(\vb{r}),   \label{eq:our_model}
\end{align}
and any effects of sample shape are fully described by the average magnetisation. Writing $\vb{H}_{\overline{\Omega}}^\text{avg}$ in terms of demagnetisation tensors, we have
\begin{align}
    \vb{H}_{\overline{\Omega}}^\text{avg}(\vb{r}) = -\mathrm{N}_{\overline{\Omega}}(\vb{r}) \vb{M}_\text{avg} = - [\mathrm{N}_\Omega(\vb{r}) - \mathrm{N}_\Lambda(\vb{r})] \vb{M}_\text{avg},   \label{eq:H_avg_demag_tensors}
\end{align}
where $\mathrm{N}_{\overline{\Omega}}$ is the demagnetisation tensor of the hollow shape $\overline{\Omega}$, while $\mathrm{N}_\Omega$ is for the full shape, and $\mathrm{N}_\Lambda$ for $\Lambda$ only. In other words, there is a scale invariant demagnetising field related to the shape by $\mathrm{N}_\Omega$, but the contribution from the average magnetisation within $\Lambda$ is included in $\vb{H}_\Lambda$, so we subtract $\mathrm{N}_\Lambda$ to avoid double counting.

The internal field of uniformly magnetised objects has been solved analytically for a range of geometries, including cylinders\cite{caciagli_exact_2018}, hemispheres\cite{durhuus_demagnetization_2025}, rectangular prisms\cite{smith_demagnetizing_2010}, tetrahedra\cite{Nielsen_tetrahedron_2019} and the general triaxial ellipsoid\cite{osborn_demagnetizing_1945,beleggia_demagnetization_2006}. For more complex geometries it can be solved numerically\cite{beleggia_computation_2003,joseph_demagnetizing_1965}. 

Our modification of the standard micromagnetic simulation amounts to evaluating two demagnetisation tensors for each cell once at the beginning of the simulation, then computing the vector average $\vb{M}_\text{avg}$ and multiplying by the 3-by-3 tensor $\mathrm{N}_{\overline{\Omega}}(\vb{r}_c)$ once per timestep. Computing $\mathrm{N}_{\overline{\Omega}}$ at each cell is an $\order{n_\text{cell}}$ operation as opposed to the $\order{n_\text{cell}^2 n_\text{copy}}$ calculation for $\mathrm{N}^\text{PBC}$, and computing $\vb{H}_\Lambda$ from $\mathrm{N}^\text{PBC}$ is an $\order{n_\text{cell}^2}$ operation to cover all cell-cell pair-interactions, while $\vb{H}_{\overline{\Omega}}^\text{avg}$ is only $\order{n_\text{cell}}$. Thus, the increase in computational expense is relatively negligible for both setup and timestepping loop.

\subsection{Relation to existing methods}

The main idea of this paper is adding $\vb{H}_{\overline{\Omega}}^\text{avg}$ to existing methods by \cref{eq:our_model,eq:H_avg_demag_tensors}. If $\vb{M}_\text{avg}=0$ or if $\Lambda = \Omega$ already, then $\vb{H}_{\overline{\Omega}}^\text{avg} = 0$ and no modification is needed, justifying the standard macrogeometry approach. That said, adding $\vb{H}_{\overline{\Omega}}^\text{avg}$ decouples the shape of $\Lambda$ from the sample shape, which gives more flexibility in defining both $\Gamma$ and the resulting macrogeometry. Also, in some cases the number of domain copies required for convergence is reduced with minimal loss of accuracy, by using $\vb{H}_{\overline{\Omega}}^\text{avg}$ for more distant regions, as demonstrated in \cref{subsec:magnetostatics}.

With PBCs out to infinity, $\Lambda$ covers all of space so that $\mathrm{N}_\Lambda = \mathrm{N}_{\mathrm{R}_3}$, yet in terms of the $\vb{H}_\text{avg}$ contribution and related shape effects, \cref{eq:our_model,eq:H_avg_demag_tensors} still apply. One caveat is that including $\Delta \vb{H}$ contributions from non-existing regions still leads to error, which is a general consideration when the entire sample is small. By translational symmetry $\mathrm{N}_{\mathrm{R}_3}(\vb{r}) = \expval{\mathrm{N}_{\mathrm{R}_3}}$ where $\expval{\mathrm{N}}$ is the volume-averaged (magnetometric) tensor, which is known to have unit trace for all shapes\cite{moskowitz_theoretical_1966} $\Tr \expval{\mathrm{N}} = 1$. By rotational symmetry, the eigenvalues are equal, so
 \begin{align}
     \mathrm{N}_{\mathbb{R}_3}(\vb{r}) = \frac{1}{3}\mqty(1 & 0 & 0 \\ 0 & 1 & 0 \\ 0 & 0 & 1),     \label{eq:N_3D}
 \end{align}
just like for the inside of a sphere or the center of a cube. Thus when simulating a spherical sample or the bulk of an isotropic macroscopic cube, infinite PBCs are already accurate, but in other cases, adding $\vb{H}_{\overline{\Omega}}^\text{avg}$ or equivalent is necessary to reintroduce sample shape in dynamical simulations.

Since $\Gamma$ is generally a tiny fraction of the sample, $\mathrm{N}_\Omega(\vb{r} = 0)$ can be used for all $\vb{r} \in \Gamma$ and similarly for $\mathrm{N}_\Lambda$ if $V_\Lambda \gg V_\Gamma$. Thus, with infinite PBCs, $\vb{H}_{\overline{\Omega}}^\text{avg}$ has no spatial dependence, making it easy to include in analytical models, or to Fourier transform for the purpose of reciprocal space methods like Ref.\ \cite{bruckner_strayfield_2021}.

Finally, we note that in uniformly magnetised, triaxial ellipsoids, the demagnetising field is uniform throughout, but it is the only finite geometry with this property\cite{maxwell_treatise_1891}. In samples of all other shapes, magnetostatic self-interaction favours a macroscopically non-uniform magnetisation, which may deviate significantly from the bulk magnetisation near the surface; especially corners and edges\cite{caciagli_exact_2018}. Additionally, magnetocrystalline anisotropy and other material properties may be different at the surface\cite{Pisane_unusual_2017}. This results in more complicated expressions for $\mathrm{N}_{\overline{\Omega}}$ in real materials, however, in using PBCs one implicitly assumes uniform magnetisation on a scale larger than $\Gamma$, so this inconsistency is a fundamental limitation of PBC models.

\section{Numerical demonstrations \label{sec:numerical_validation}}

\subsection{System of interest \label{subsec:system_of_interest}}

\begin{figure}
    \centering
    \includegraphics[width=0.7\linewidth]{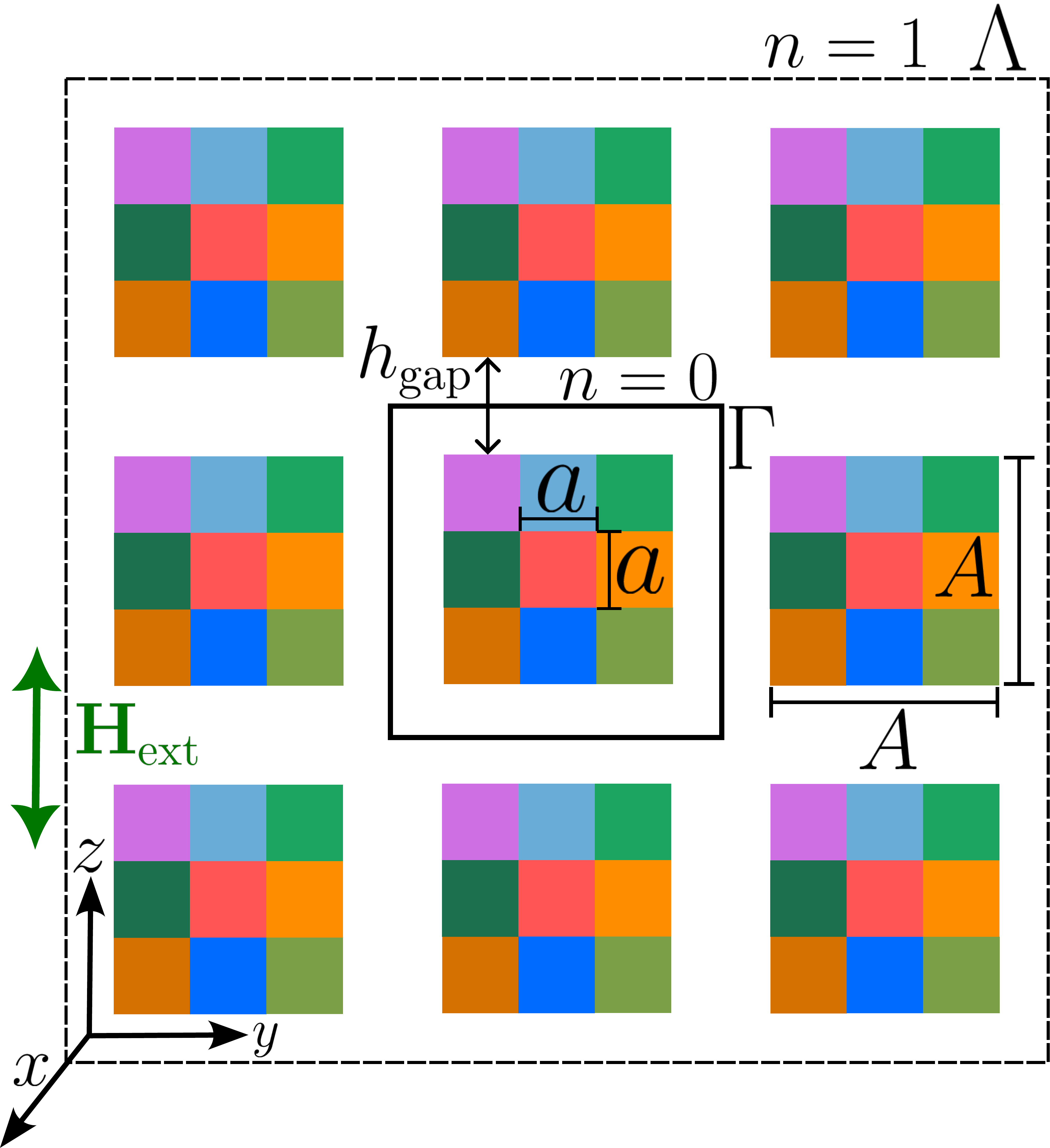}
    \caption{Illustration of the simulated system, i.e.\ a collection of magnetic grains separated by non-magnetic material and subject to an oscillating external field, $\vb{H}_\text{ext}$. The example shows a 2D slice of the macrogeometry with one layer of domain copies and $3\times3\times3$ micromagnetic cells in each grain. Equivalent cells in different domain copies are the same color.}
    \label{fig:macrogeometry}
\end{figure}

To validate and demonstrate the presented formalism, we consider a sample of magnetic grains uniformly distributed in 3D as illustrated in \cref{fig:macrogeometry}. Each grain is a cube of sidelength $A$ consisting of $n_\text{side} \times n_\text{side} \times n_\text{side}$ micromagnetic cells which are themselves uniformly magnetised cubes of sidelength $a$ and magnetisation $M_s$.
Neighbouring grains are separated by an $h_\text{gap}$ thick layer of non-magnetic material on all sides. The simulated domain $\Gamma$ is a single grain and half the surrounding vacuum layer, i.e.\ a cube of sidelength $A + h_\text{gap}$. Thus, the PBCs entail that every grain has the same magnetisation distribution. The macrogeometry $\Lambda$ consists of $\Gamma$ itself and $n$ layers of domain copies as shown in \cref{fig:macrogeometry}. Hence, there are $(2n+1)^{n_\text{dim}}$ grains in $\Lambda$ where $n_\text{dim} = 1,2,3$ for 1D, 2D and 3D periodic boundary conditions respectively. The macroscopic sample is itself a rectangular prism (box) centered on the origin with sidelengths $(1, 1, p) \cdot  A_\text{sample}$, i.e.\ equal along $x$ and $y$, and $p A_\text{sample}$ along $z$ so $p$ is the aspect ratio. For both $\mathrm{N}_\Omega, \mathrm{N}_\Lambda$ and  the single-cell demagnetisation tensors $\mathrm{N}_c$ we use the exact solution for uniformly magnetised prisms\cite{smith_demagnetizing_2010}, of which the cube is a special case.

For the purpose of micromagnetism, we further include a uniaxial anisotropy along $x$ of anisotropy constant $K$ and a uniform sinusoidal field along $z$ of amplitude $H$ and frequency $f$. Then, the equation of motion for the magnetisation of cell $c$ is
\begin{align}
    \dv{t} \vb{m}_c = - \gamma \vb{m}_c \times \vb{H}^\text{eff}_c - \eta \vb{m}_c \times (\vb{m}_c \times \vb{H}^\text{eff}_c),
\end{align}
where $\vb{m}_c = \vb{M}_c / M_s$ is the normalised magnetisation vector, $\gamma$ is the gyromagnetic ratio, $\eta = \frac{\alpha}{1 + \alpha^2} \gamma$
is the damping constant, where $\alpha$ is the dimensionless Gilbert damping, and the effective field at time $t$ is
\begin{align}
    \vb{H}^\text{eff}_c &= \frac{2A_\text{ex}}{\mu_0 M_s} \nabla^2 \vb{m}_c - \frac{2K}{\mu_0 M_s} (\vb{\hat{x}} \vdot \vb{m}_c) \vb{\hat{x}} 
    \notag \\
    &\quad H \sin(2\pi f t) \vb{\hat{z}} + \vb{H}^\text{demag}_c,
\end{align} 
with the exchange stiffness $A_\text{ex}$ quantifying the ferromagnetic exchange coupling between nearby cells. We use a constant timestep of $\Delta t = 0.5 \: \mathrm{ps}$ for all simulations.
The studied parameters are given in \cref{tab:parameters}.

\begin{table}[tbh]
    \centering
    \begin{tabular}{|c|c|c|c|}
        \hline
        Symbol & Description & Unit & Value(s)  \\
        \hline
        $\mu_0 H $ & Field amplitude & $\mathrm{T}$ & $0.5$ \\
        \hline
        $f$ & Field frequency & $\mathrm{MHz}$ & $100$ \\
        \hline
        $\mu_0 M_s$ & Saturation magnetisation & $\mathrm{T}$ & $1.5$ \\
        \hline
        $K$ & Anisotropy constant & $\mathrm{kJ / m^{3}}$ & $8$ \\
        \hline
        $A_\text{ex}$ & Exchange stiffness & $\mathrm{J/nm}$ & $100$ \\
        \hline
        $\alpha$ & Gilbert damping
        & - & 1 \\
        \hline
        $\gamma$ & Gyromagnetic ratio & $\mathrm{GHz/T}$ & $176$ \\
        \hline
        $a$ & Sidelength of cell & $\mathrm{nm}$ & $7$ \\
        \hline
        $n_\text{side}$ & $n_\text{side}^3$ cells in $\Gamma$ & $-$ & $10, 15$ \\
        \hline
        $p$ & Aspect ratio of sample & $-$ & 
        $0..2$
        \\
        \hline
        $h_\text{gap}$ & Intergrain gap & $\mathrm{nm}$ & $0, 1, 10, 100, 1000$ \\
        \hline
        $\Delta t$ & Timestep & $\mathrm{ps}$ & 0.5 \\
        \hline
    \end{tabular}
    \caption{Simulated parameters. 
    The simulated domain contains an $n_\text{side} \times n_\text{side} \times n_\text{side}$ grid of micromagnetic cubes with uniaxial anisotropy along $x$ and a uniform, sinusoidal magnetic field applied along $z$. }
    \label{tab:parameters}
\end{table}

\subsection{Magnetostatics \label{subsec:magnetostatics}}

\begin{figure}[h]
    \centering
    \includegraphics[width=0.9\linewidth]{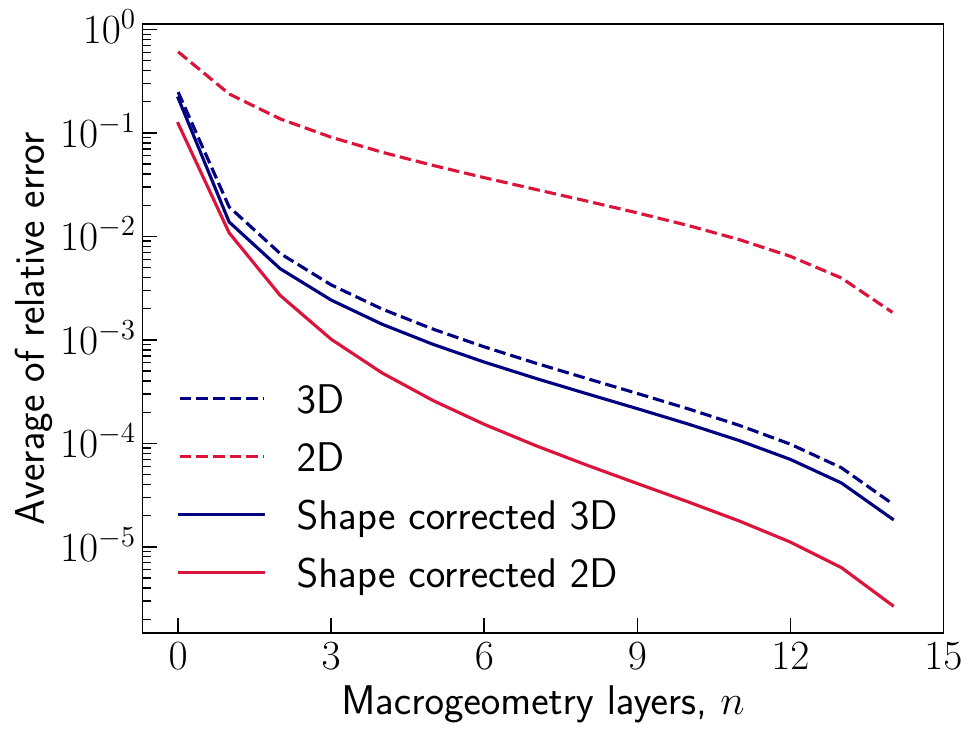}
    \caption{Convergence rate vs.\ number of macrogeometry layers for the system in \cref{fig:macrogeometry} with infinite PBS in 2D (periodic along $x,y$) or 3D (periodic along $x,y,z$). Results are shown with and without the shape-correction field \cref{eq:H_avg_demag_tensors,eq:our_model}. For each method, errors are computed relative to the $n=15$ demagnetisation field calculated with the same method and averaged over 1000 positions.}
    \label{fig:magstatic_convergence}
\end{figure}

We consider a system as described in \cref{subsec:system_of_interest} with $n_\text{side} = 10$, $h_\text{gap} = 1\:\mathrm{nm}$ and 2D or 3D PBCs out to infinity.
For the 3D case, $\mathrm{N}_\Omega = \mathrm{N}_{\mathbb{R}_3}$ is given in \cref{eq:N_3D} while with 2D infinite PBCs the sample is equivalent to an infinitesimally thin plate ($p=0$), so $\mathrm{N}_\Omega = \mathrm{N}^\text{plate}$ where $N^\text{plate}_{zz} = 1$ and all other components are 0 \cite{blundell_magnetism_2001}. 

To test how quickly the demagnetisation field converges as function of macrogeometry layers $n$, we assign a randomly distributed magnetisation to the simulated grain in $\Gamma$. Specifically, the polar and azimuthal angle of $\vb{m}_c$ are $\theta_0 + \theta_c$ and $\phi_0 + \phi_c$ where $\theta_0, \phi_0$ are uniformly random while $\theta_c, \phi_c$ are drawn from normal distributions with standard deviations of $\pi/3$ and $2\pi/3$ respectively. Because $\theta_0, \phi_0$ are the same for all cells, we get a significant average magnetisation. The same magnetisation is used for all $n$ values and for each of four cases: 2D and 3D PBCs with/without the shape correction field \cref{eq:H_avg_demag_tensors}. In each case, we use the field with $n=15$ macrogeometry layers as reference $\vb{H}_\text{ref}$ and define the relative error at position $\vb{r}$ by
\begin{align}
    H_\text{error}(\vb{r}) = \frac{\abs{\vb{H}_\text{demag}(\vb{r}) - \vb{H}_\text{ref}(\vb{r})}}{\abs{\vb{H}_\text{ref}(\vb{r})}}.
\end{align}
We average this error over 1000 points randomly distributed within the magnetised part of $\Gamma$ (same 1000 points used throughout).

The convergence test is shown in \cref{fig:magstatic_convergence}. We emphasise that the figure only shows convergence rate, not accuracy, but the mean error between the two reference fields ($n=15$ with/without shape correction) is only $0.01\: \%$ for 3D and $2.6\%$ for 2D PBCs, so within the convergence error, the methods are mutually consistent.
Shape corrections consistently improve convergence, but the change is small in 3D, because the cubical macrogeometry has the correct shape for all $n$. Unlike $\mathbb{R}_3$, a finite magnetised cube has a slightly non-uniform internal field, so the sole effect of shape-correcting is to homogenise the field from average magnetisation, removing a finite-size artifact. 

Conversely, with 2D PBCs, the macrogeometry only reflects the true aspect ratio in the limit of $n\rightarrow \infty$, which results in very poor convergence. Using the computationally inexpensive shape correction, the relative error is almost an order of magnitude lower at $n=0$ and the gap grows with increasing $n$. 
This indicates that the vast majority of the convergence error is due to the incorrect aspect ratio, so shape-correction is essential with lower-dimensional PBCs to infinity. With shape-correction, the 2D case converges significantly faster than 3D. The reason is that the net field from a given macrogeometry layer is generally smaller when $\Gamma$ is only copied in some directions. Hence, the contribution from non-uniform magnetisation is shorter-ranged with 1D and 2D PBCs than 3D, and thus easier to converge. 

\subsection{Micromagnetics \label{subsec:micromagnetics}}

\begin{figure}[htb]
    \centering
    \includegraphics[width=0.75\linewidth]{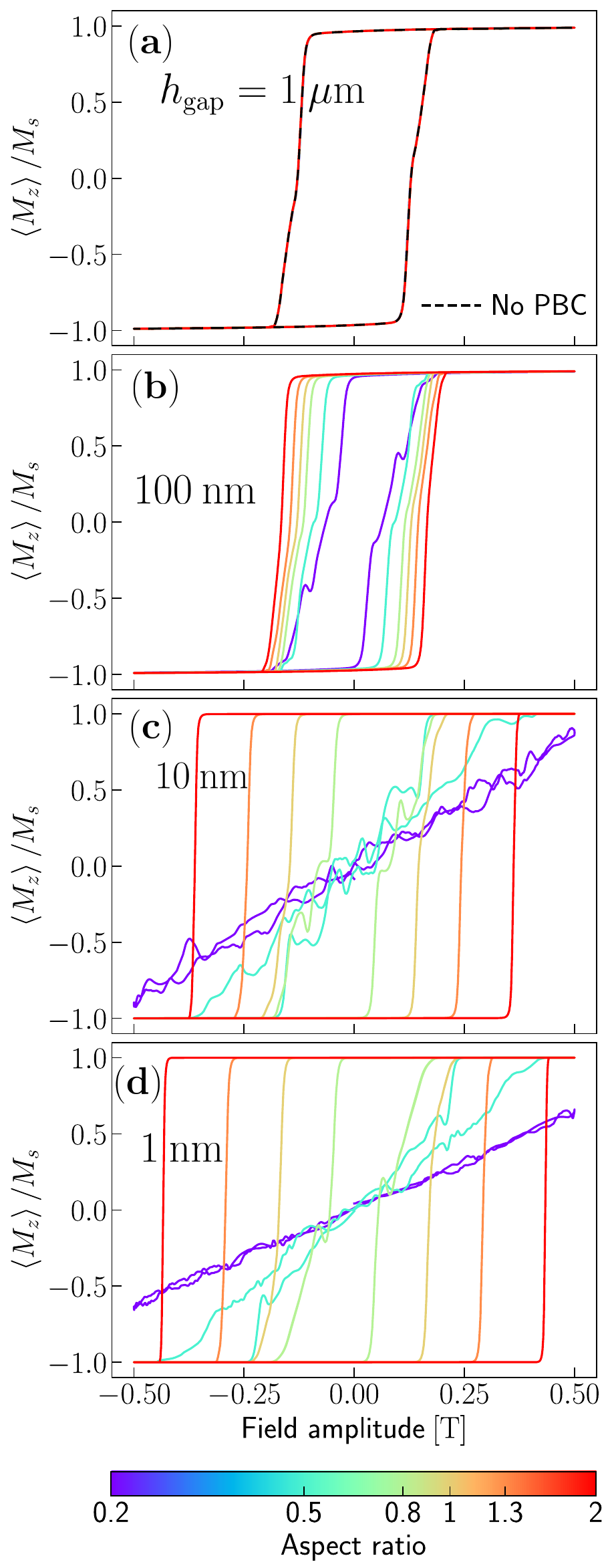}
    \caption{Normalised magnetisation vs.\ applied field (hysteresis curves) for different aspect ratios (colorcoded) and four different intergrain spacings (different plots). Lines and numbers on colorbar show the 6 aspect ratios used. \textbf{(a)} also has a dashed black line showing the simulated hysteresis curve for a lone grain ($n=0$ and no shape correction).}
    \label{fig:hysteresis_vs_shape}
\end{figure}

We again consider a system of uniformly distributed, magnetised grains as described in \cref{subsec:system_of_interest} and sketched in \cref{fig:macrogeometry}. The sample is a macroscopic rectangular prism; we set $n_\text{side} = 15$, use 2 macrogeometry layers i.e.\ 25 domain copies, and vary aspect ratio from $p=0.2$ to $p=2$ for 5 different intergrain gaps $h_\text{gap}$. The $h_\text{gap}$ values and other relevant parameters are given in \cref{tab:parameters}. We set $A_\text{sample} = 1\:\mathrm{m}$, so $\mathrm{N}_\Omega(\vb{r}) \approx \mathrm{N}_\Omega(\vb{0})$ throughout $\Gamma$.

The simulated sample is a magnetically soft, high $M_s$ composite which corresponds well to e.g.\ a high-frequency inductor core. For high-frequency applications, composites are used in place of bulk magnets despite reduced permeability, because having the conductive material in separate, microscopic grains massively reduces the heating from induced eddy currents. Similar systems have been modelled by macroscopic electromagnetism\cite{Wier_susceptibility_2022,Arzbacher_tomography_2015} and micromagnetism\cite{bruckner_strayfield_2021,ducevic_micromagnetic_2025}. 

The computed hysteresis curves are shown in \cref{fig:hysteresis_vs_shape}. When $h_\text{gap} = 1\:\mathrm{\mu m}$ (\cref{fig:hysteresis_vs_shape}a), there is no observable shape effect as all curves coincide with that of an isolated magnetic grain. The reason is that $M_\text{avg} \leq \phi M_s$ where $\phi$ is the volume fraction of magnetic material, and since the sidelength of a grain is only $A = n_\text{side} a = 105\:\mathrm{nm}$,
\begin{align}
    \phi = \frac{A^3}{(A+h_\text{gap})^3} = 0.086\: \%,
\end{align}
so the field from intergrain coupling is a negligible fraction of the internal field. In other words, the grains are far enough apart to be essentially non-interacting.

The next question is if the non-interacting curve is reasonable? If the external field $\vb{H}_\text{ext}$ was low frequency, the Stoner-Wohlfarth model of uniformly magnetised grains would serve as an upper bound on the coercive field. Since the anisotropy axis is perpendicular to $\vb{H}_\text{ext}$ no coercivity is expected and even with parallel anisotropy the coercive field would be $H_\textrm{coercive} \leq \frac{2K}{M_s} = 13\:\mathrm{mT}$, i.e. about a factor 10 smaller than observed. The explanation is that the external field is oscillating too quickly for the multi-domain magnetisation reversal to keep up, so the hysteresis curve is dominated by dynamical effects. For a related system, Ref.\ \cite{ducevic_micromagnetic_2025} found the dynamical hysteresis power was well described by the phenomenological Steinmetz equation\cite{reinert_steinmetz_2001} $P_\text{dyn} \sim H^\beta f^\delta$ where $\beta, \delta$ are fitting parameters found to be $\beta=1.93$ and $\delta = 1.71$ for the system of Ref.\ \cite{ducevic_micromagnetic_2025}. Because the area of the hysteresis curve $A_\text{hyst}$ is also the energy absorbed per field cycle, we have $P_\text{dyn} = f A_\text{hyst}$, so the coercive field for a square loop must obey
\begin{align}
    H_\text{coercive} \sim H^\beta f^{\delta - 1},
\end{align}
which explains the rather large coercivities observed.

\begin{figure}[htb]
    \centering
    \includegraphics[width=\linewidth]{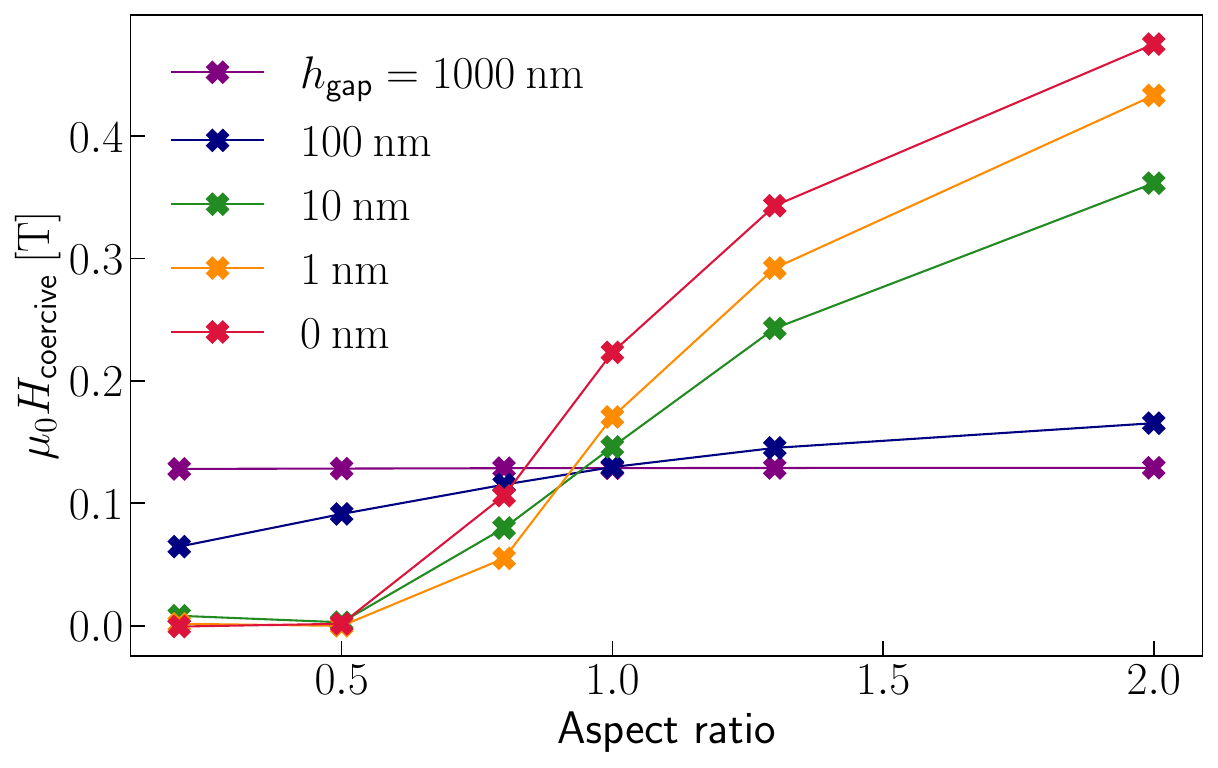}
    \caption{Coercive field vs.\ aspect ratio and intergrain gap, with periodic exchange included for $h_\text{gap} = 0$.}
    \label{fig:coercivity_vs_shape}
\end{figure}

For lower $h_\text{gap}$ (\cref{fig:hysteresis_vs_shape}b-d) sample shape becomes significant. Elongating along $z$ produces easy-axis shape anisotropy which favours alignment with $\vb{H}_\text{ext}$ but also increases the cost of reversal. Indeed we observe broader and more square loops the higher $p$ is. Conversely, $p<1$ creates a hard-axis along $z$, thus favouring in-plane vortex states where the $z$ component increases gradually with applied field. This is qualitatively consistent with the Stoner-Wohlfarth model, but we also observe non-monotonous magnetisation changes for $p < 1$ and an opening of the curves at high field despite zero remanence for the lightblue curves in \cref{fig:coercivity_vs_shape}c-d ($p = 0.5$). Both effects we attribute to the high-frequency excitation of multi-domain states.

Finally, in \cref{fig:coercivity_vs_shape} we show the coercivity vs.\ field for the same data as \cref{fig:hysteresis_vs_shape} plus a simulation sequence with $h_\text{gap} = 0$, i.e.\ a continuous magnet. For $h_\text{gap} = 0$ we included periodic exchange, so for instance the cells on the left face of $\Gamma$ are exchange coupled to those on the right face. This made no qualitative changes, but consistently increased coercivity as the stronger exchange coupling stabilises the fully aligned state.

\section{Conclusion and outlook \label{sec:conclusion}}

We developed a simple technique for including the effect of sample shape in micromagnetism using a shape correction tensor which only acts on the average magnetisation. The resulting shape correction field is computationally cheap, can readily be added to existing PBC schemes and we proved formally that the method converges even in 3D whenever magnetisation is macroscopically uniform; as is implicitly assumed in all PBC methods.
One still needs a conventional PBC scheme for the demagnetisation field of nearby regions to account for locally non-uniform magnetisation, but the shape correction reintroduces shape effects to schemes with infinite PBCs, adds flexibility to macrogeometry methods. Even when the macrogeometry shape reflects the sample, there is a modest improvement in convergence rate as finite-size artifacts are removed. 

We applied the shape corrected macrogeometry method to a soft magnetic composite in a $100\: \mathrm{MHz}$ applied field. The hysteresis curves in this high-frequency regime are dominated by dynamical effects, but the impact of sample aspect ratio is mostly the same as at low-frequency e.g.\ an increase in coercivity the more the sample is elongated along the field axis.

One of the main, remaining challenges in simulating mascroscopic samples with micromagnetics is the study of surface effects, as the local magnetisation may differ greatly from the bulk value. With the present formalism one can solve for the bulk hysteresis curve, then use the resulting $\vb{M}_\text{avg}(t)$ and combined vacuum/PBC boundary conditions to get the surface demagnetisation field in the same external field. Of course the surface magnetisation affects $\vb{M}_\text{avg}$, so the method is not fully self-consistent. We leave further refinement and validation of this idea to future work.

\section*{Data statement}
All data presented in this work are available from Ref. 



\appendix

\section{Proof that average magnetisation determines shape anisotropy of a macroscopic sample \label{appsec:3D_PBC_proof}}

Let $\Gamma$ be the simulated domain and let there be infinite copies of $\Gamma$, denoted $\Gamma_i$, which tile all of 3D space, $\mathbb{R}_3$, with no overlap. Furthermore, let $c$ index the cells within $\Gamma$ with magnetisations $\vb{M}_c$ and positions $\vb{r}_c$ relative to the center of $\Gamma$. Here we prove that if the magnetic sample is large enough, the shape-dependent part of the demagnetising field within $\Gamma$ is fully determined by the average magnetisation,
\begin{align}
    \vb{M}_\text{avg} = \frac{1}{V_\Gamma} \int_\Gamma \vb{M} \dd \vb{r},   \label{eq:M_avg}
\end{align}
where $V_\Gamma$ is the volume of $\Gamma$. By the PBCs, $\vb{M}_\text{avg}$ is also the average magnetisation of each $\Gamma_i$ and of the entire sample.

\subsection{Decomposing magnetic field \label{appsec:decomposing_magnetic_field}}

The field from domain $\Gamma_i$ at position $\vb{r}$ is given by
\begin{align}
    \vb{H}_i(\vb{r}) = \frac{1}{4\pi} \int_{\Gamma} \frac{3 [\vb{M}(\vb{r}') \vdot \Delta \vb{\hat{r}}] \Delta \vb{\hat{r}} - \vb{M}(\vb{r}')}{\Delta r^3} \dd \vb{r'},  \label{eq:H_dip}
\end{align}
where $\vb{r}'$ is position relative to the center of $\Gamma_i$ denoted $\vb{r}_i$, while $\Delta \vb{r} = \vb{r} - \vb{r}' - \vb{r}_i$ and $\Delta \vb{\hat{r}} = \frac{\Delta \vb{r}}{\Delta r}$ with $\Delta r = \abs{\Delta \vb{r}}$. Because $\vb{H}_i$ is linear in magnetisation, we can separate the average magnetisation from the rest by writing
\begin{align}
    \vb{H}_i = \vb{H}_i^\text{avg} + \Delta \vb{H}_i    \label{eq:decomposed_field}
\end{align}
where $\vb{H}_i^\text{avg}$ and $\Delta \vb{H}_i$ are also given by \cref{eq:H_dip} except that 
$\vb{M} \xrightarrow{} \vb{M}_\text{avg}$ in the former and $\vb{M} \xrightarrow{} \Delta \vb{M} = \vb{M} - \vb{M}_\text{avg}$ in the latter. Crucially, by \cref{eq:M_avg}, $\int_\Gamma \Delta \vb{M} \dd \vb{r} = 0$.

\subsection{Rewriting the field from non-uniform magnetisation \label{appsec:rewriting_Delta_H}}

The magnetic field resulting from non-uniform magnetisation is given by $\Delta \vb{H}_i$ from \cref{eq:decomposed_field}, which we may write as
\begin{align}
    \Delta \vb{H}_i = -\frac{1}{4\pi} \int_{\Gamma} \mathrm{N}^\text{dip}(\Delta \vb{r}) \Delta \vb{M}(\vb{r}') \dd\vb{r}',   \label{eq:H_ijk}
\end{align}
where
\begin{align}
    N_{\alpha\beta}^\text{dip}(\Delta \vb{r}) = \frac{1}{\Delta r^3} \delta_{\alpha \beta} - \frac{3 \Delta r_\beta \Delta r_\alpha}{\Delta r^5}    \label{eq:N_dip}
\end{align}
is the demagnetisation tensor for a dipole, $\alpha, \beta = x,y,z$ while $\Delta r_x = \Delta x$ and similarly for $\Delta r_y, \Delta r_z$.
Without loss of generality we set the origin at the evaluation point so $\vb{r} = 0$. We note that $\mathrm{N}_\text{dip}(-\vb{r}-\vb{r}_i) = \mathrm{N}_\text{dip}(\vb{r} + \vb{r}_i)$ and Taylor expand around $\vb{r}'=0$. The general formula in Cartesian coordinates is
\begin{align}
    &N_{\alpha\beta}^\text{dip}(\vb{r}' + \vb{r}_i)
     \notag \\
    &\hspace{0.3cm}=\hspace{-0.3cm}\sum_{n_x,n_y,n_z=0}^\infty \hspace{-0.3cm}\eval{\partial_{x'}^{n_x}\partial_{y'}^{n_y}\partial_{z'}^{n_z} N_{\alpha\beta}^\text{dip}(\vb{r}'+\vb{r}_i)}_{\vb{r}' = 0} \frac{x'^{n_x} y'^{n_y} z'^{n_z}}{n_x!n_y!n_z!} 
    \notag \\
    &\hspace{0.3cm}= \hspace{-0.3cm}\sum_{n_x,n_y,n_z=0}^\infty \hspace{-0.3cm} \partial_{x'}^{n_x}\partial_{y'}^{n_y}\partial_{z'}^{n_z} \eval{N_{\alpha\beta}^\text{dip}(\vb{r}' + \vb{r}_i)}_{\vb{r}' = 0} \hspace{-0.2cm} r'^{n_x+n_y+n_z} g(\vb{\hat{r}'}), \label{eq:Taylor_series}
\end{align}
where $g$ is a dimensionless function of the direction of $\vb{r}'$. Next, we invoke the identity
\begin{align}
    \pdv{r_\alpha'} \frac{r_\alpha'^{n_1}}{\abs{\vb{r}' + \vb{r}_i}^{n_2}} = \frac{n_1 r_\alpha'^{n_1 - 1}}{\abs{\vb{r}' + \vb{r}_i}^{n_2}} - \frac{n_2 r_\alpha'^{n_1+1}}{\abs{\vb{r}' + \vb{r}_i}^{n_2+2}}.  \label{eq:Taylor_identity}
\end{align}
This expression can in principle be used iteratively to perform every differential operation, yielding a terribly complicated expression. However, we note that only terms of the form $a/r_i^n$ where $a$ is independent of $\vb{r}'$ are left after letting $\vb{r}' \xrightarrow{} 0$ and that applying \cref{eq:Taylor_identity} can never decrease $n$. To leading order in $1/r_i$, clearly $\mathrm{N}^\text{dip} \sim 1/r_i^3$ and $N^\text{dip}$ has units of inverse volume. It follows from \cref{eq:Taylor_series,eq:Taylor_identity} and dimensional analysis, that the Taylor series has to take the form
\begin{align}
    \mathrm{N}^\text{dip}(\vb{r}' + \vb{r}_i) = \sum_{n=0}^\infty \frac{r'^n}{r_i^{n+3}} \mathrm{G}_n(\vb{\hat{r}'}, \vb{\hat{r}}_i),   \label{eq:N_dip_expansion}
\end{align}
where $n=n_x+n_y+n_z$, and $\mathrm{G}_{n}(\vb{\hat{r}'}, \vb{\hat{r}}_i)$ is a dimensionless, 3-by-3 tensor function of the directions of $\vb{r'}$ and $\vb{r}_i$. Comparing with \cref{eq:Taylor_series} we see that $\mathrm{G}_0$ is independent of $\vb{r}'$, so
inserting in \cref{eq:H_ijk} and using that $\int_\Gamma \Delta \vb{M} \dd \vb{r} = 0$ by definition we find that the $n=0$ term is zero. What remains is
\begin{align}
     \Delta \vb{H}_i = \frac{1}{4\pi}\sum_{n=1}^\infty \frac{1}{r_i^{n+3}} \int_\Gamma r'^n \mathrm{G}_n \Delta \vb{M} \dd\vb{r'}  \label{eq:Delta_H}
\end{align}
The fact that the leading order term is $\sim r_i^{-4}$ is crucial.

\subsection{Proof of convergence for higher order contributions}

In the remainder of this proof, we show that the combined $\Delta \vb{H}$ field from all domain copies beyond a certain distance is negligible. To this end, let $\Lambda$ be the union of $\Gamma$ and its $N$ closest copies and let $\neg \Lambda = \mathbb{R}_3 - \Lambda$ be the rest of 3D space. We are interested in an upper bound for the field magnitude
\begin{align*}
    \Delta H_{\neg \Lambda} = \frac{1}{4\pi} \abs{\sum_{\substack{i \\ \Gamma_i \notin \Lambda}} \sum_{n=1}^\infty \frac{1}{r_i^{n+3}} \int_{\Gamma} r'^n \mathrm{G}_n \Delta \vb{M} \dd \vb{r'}}.
\end{align*}
By the triangle inequality $\abs{\vb{v}_1 + \vb{v}_2} \leq \abs{\vb{v}_1} + \abs{\vb{v}_2}$ for any vectors $\vb{v}_1, \vb{v}_2$, hence 
\begin{align*}
    \Delta H_{\neg \Lambda} &\leq \frac{1}{4\pi}\sum_{\substack{i \\ \Gamma_i \notin \Lambda}} \sum_{n=1}^\infty \frac{1}{r_i^{n+3}} \int_{\Gamma} r'^n \abs{\mathrm{G}_n \Delta \vb{M}} \dd \vb{r'}
    \\
    &\leq \frac{1}{4\pi} \sum_{n=1}^\infty \norm{\mathrm{G}_n \Delta \vb{M}}_\text{max} R^n_\Gamma \sum_{\substack{i \\ \Gamma_i \notin \Lambda}}\frac{V_\Gamma}{r_i^{n+3}},
\end{align*}
where $R_\Gamma$ is the largest distance from the center of $\Gamma$ to its surface and $\norm{\mathrm{G}_n \Delta \vb{M}}_\text{max}$ is the largest value the magnitude of the vector takes within $\Gamma$. Since we replaced the integrand with its maximal value, the $\vb{r}'$ integral simply becomes $V_\Gamma$.
Because $V_\Gamma$ is the volume of every domain copy, the inner sum can be interpreted as the volume of the entire infinite space $\neg \Lambda$, with each copy $\Gamma_i$ weighted by $\frac{1}{r_i^{n+3}}$. It follows that we can introduce an integral bound by
\begin{align*}
    \Delta H_{\neg \Lambda} &\leq \frac{1}{4\pi} \sum_{n=1}^\infty \norm{\mathrm{G}_n \Delta \vb{M}}_\text{max} R^n_\Gamma \int_{\neg \Lambda} \frac{1}{(r - R_\Gamma)^{n+3}} \dd \vb{r},
\end{align*}
To see that this is an upper bound, one must realise that for every region $\Gamma_i$ not in $\Lambda$ we have $r - R_\Gamma \leq r_i$ hence
the integrand is $\geq \frac{1}{r_i^{n+3}}$. Now let $R_\Lambda$ be the smallest distance from the center of $\Lambda$ to its surface. A sphere with this radius is fully contained within $\Lambda$ and the integrand is non-negative, so integrating from $R_\Lambda$ to $\infty$ necessarily gives a larger value than integrating over $\neg \Lambda$, hence
\begin{align*}
    \Delta H_{\neg \Lambda} &\leq \frac{1}{4\pi} \sum_{n=1}^\infty \norm{\mathrm{G}_n \Delta \vb{M}}_\text{max} R^n_\Gamma \int_{r \geq R_\Lambda} \frac{1}{(r - R_\Gamma)^{n+3}} \dd \vb{r}
    \notag\\
\end{align*}
Because of rotational symmetry, the integral is relatively straightforward to evaluate in spherical coordinates, yielding  
\begin{align}
    \Delta H_{\neg \Lambda} &\leq \sum_{n=1}^\infty \norm{\mathrm{G}_n \Delta \vb{M}}_\text{max} R^n_\Gamma \int_{R_\Lambda}^\infty \frac{r^2 \; \dd r}{(r-R_\Gamma)^{n+3}}
    \label{eq:key_integral}\\
    &= \sum_{n=1}^\infty \norm{\mathrm{G}_n \Delta \vb{M}}_\text{max} \frac{R_\Gamma^n}{n} \frac{R_\Lambda^2}{(R_\Lambda - R_\Gamma)^{n+2}} 
    \notag\\
    &\: \times \left(1 - \frac{(2n + 4) R_\Gamma}{(n^2+3n+2)R_\Lambda} + 2 \frac{R_\Gamma^2}{(n^2+3n+2)R_\Lambda^2} \right).
    \notag 
\end{align}
We use that $R_\Gamma$ is fixed while $R_\Lambda$ can be made arbitrarily large for the purposes of this proof so we can always choose $\frac{R_\Gamma}{R_\Lambda} \leq 1$, hence
\begin{align}
    \Delta H_{\neq \Lambda}  &\leq \sum_{n=1}^\infty \norm{\mathrm{G}_n \Delta \vb{M}}_\text{max} \frac{n + 1}{n^2 + 3n + 2} \frac{R_\Lambda^2 R_\Gamma^n}{(R_\Lambda - R_\Gamma)^{n+2}}, \notag \\
    &\leq \sum_{n=1}^\infty \norm{\mathrm{G}_n \Delta \vb{M}}_\text{max} \frac{R_\Lambda^2 R_\Gamma^n}{(R_\Lambda - R_\Gamma)^{n+2}}.     \label{eq:H_final_sum}
\end{align}
When $N \xrightarrow{} \infty$ the region $\Lambda$ grows indefinitely in all directions, so $R_\Lambda \xrightarrow{} \infty$, while the remaining variables are unchanged, hence the sum goes to zero term by term as $\left(\frac{R_\Gamma}{R_\Lambda}\right)^{n}$. That is
\begin{align}
    \lim_{N \xrightarrow{} \infty} \Delta H_{\neg \Lambda} = \lim_{R_\Lambda \xrightarrow{} \infty} \Delta H_{\neg \Lambda} = 0.  \label{eq:H_limit_proven}
\end{align}
We note that for $n=0$, the integral in \cref{eq:key_integral} diverges, as the antiderivative has a logarithmic term, so the proof does not apply to $\vb{H}_i^\text{avg}$. 
This is as it should be, since the internal demagnetisation field from a uniformly magnetised domain is known to be scale-invariant.

\subsection{Exchanging sum and limit}
Technically to take the $R_\Lambda \xrightarrow{} \infty$ limit term-by-term in \cref{eq:H_final_sum}, one first has to prove the infinite sum has a finite value, so we note that when $R_\Lambda \geq 2R_\Gamma$
\begin{align*}
     &\sum_{n=1}^\infty \norm{\mathrm{G}_n \Delta \vb{M}}_\text{max} \frac{n + 1}{n^2 + 3n + 2} \frac{R_\Lambda^2 R_\Gamma^n}{(R_\Lambda - R_\Gamma)^{n+2}} 
     \notag \\
     &= \sum_{n=1}^\infty A_n \norm{\mathrm{G}_n \Delta \vb{M}}_\text{max}
    <  \sum_{n=1}^\infty \norm{\mathrm{G}_n \Delta \vb{M}}_\text{max},
\end{align*}
since $A_n \leq A_1 \leq 1$ with $A_1 = 1$ for $R_\Lambda = 2R_\Gamma$. Because $R_\Gamma$ is fixed while $R_\Lambda$ can be made arbitrarily large, the assumption $R_\Lambda \geq 2R_\Gamma$ causes no loss of generality. Next, we use that
\begin{align}
    \sum_{n=1}^\infty \norm{\mathrm{G}_n \Delta \vb{M}}_\text{max} \leq M_s \sum \max_{\alpha \beta \vb{r}'}\abs{G_{\alpha\beta n}(\vb{\hat{r}}', \vb{\hat{r}}_i)},
\end{align}
where $M_s$ is the saturation magnetisation. Thus, it is sufficient to prove each component of $\mathrm{G}_n$ is absolutely convergent for any direction vectors, $\vb{\hat{r}}', \vb{\hat{r}}_i$. 

Since the dipole tensor components $N^\text{dip}_{\alpha\beta}(\vb{r}' + \vb{r}_i)$ are finite everywhere except $\vb{r}' = - \vb{r}_i$ (cf.\ \cref{eq:N_dip}), it follows from \cref{eq:N_dip_expansion} that the series' $\frac{r'^n}{r_i^n} G_{\alpha\beta n}$ converges. Hence, by the Cauchy root test\cite{Knopp_infinite_1990},
\begin{align*}
    \lim_{n\xrightarrow{} \infty} \sqrt[n]{\frac{r'^n}{r_i^n}\abs{G_{\alpha \beta n}}} \leq 1 \Rightarrow \lim_{n\xrightarrow{} \infty} \sqrt[n]{\abs{G_{\alpha \beta n}}} \leq \frac{r_i}{r'}.
\end{align*}
Because $\mathrm{G}_n$ does not depend on the magnitudes $r',r_i$, we can freely set $r_i/r' < 1$, so it follows from the root test that the series' $G_{\alpha \beta n}$ are absolutely convergent. Hence the final sum in \cref{eq:H_final_sum} is indeed finite so \cref{eq:H_limit_proven} is true.

\subsection{Necessity of macroscopically uniform magnetisation}

With PBCs the sample magnetisation is modelled as uniform on scales much larger than the characteristic length of the simulated domain, $R_\Gamma$. Here we show the necessity of this macroscopic uniformity for the preceding proof to hold.

Consider the magnetisation
\begin{align}
    \vb{M} = M \Theta(r \geq R_1) \Theta(r \leq R_2) \vb{\hat{z}} \begin{cases}
        1 & \theta < \frac{\pi}{4} \lor \theta > \frac{3\pi}{4} \\ -1 & \theta \in \left[\frac{\pi}{4}, \frac{3\pi}{4}\right],
    \end{cases} \label{eq:M_pathological}
\end{align}
where $\Theta$ is the Heaviside step function, $\lor$ a logical "or", $\theta$ is the polar angle in spherical coordinates and $R_2 > R_1$. The magnetisation is zero outside a spherical shell from $r = R_1$ to $r = R_2$. Within said shell, it points downwards for the angular interval $\frac{\pi}{4} \leq \theta \leq \frac{3\pi}{4}$ and upwards otherwise. The average magnetisation across all space is zero, yet there is evidently a consistent magnetic order. The field generated at the origin, i.e.\ the center of the spherical shell, is
\begin{align}
    \vb{H} &= \frac{1}{4\pi} \int_{R_2 \geq r \geq R_1} \frac{3 (\vb{M} \vdot \vb{\hat{r}}) \vb{\hat{r}} - \vb{M}}{r^3} \dd\vb{r}
    \notag\\
    &= \frac{M}{4\pi} \vb{\hat{z}} \int_{R_1}^{R_2} \frac{\dd r}{r} \left(\int_0^{\frac{\pi}{4}} \dd \theta + \int_{\frac{3\pi}{4}}^{\pi} \dd\theta - \int_{\frac{\pi}{4}}^{\frac{3\pi}{4}} \dd \theta \right)
    \notag\\
    &\qquad \times (3 \cos^2 \theta - 1) \sin \theta
    \notag\\
    &= \frac{\sqrt{2} M}{4\pi} \ln \left(\frac{R_2}{R_1}\right) \vb{\hat{z}}.   \label{eq:M_pathological_calc}
\end{align}
In this case, no matter how large $R_1$ is, the field strength diverges when $R_2 \xrightarrow{} \infty$. The same logarithmic divergence of the central field is known from Halbach arrays, where the magnetisation rotates continuously\cite{Halbach_Design_1980,Bjork_analysis_2010,Bjork_efficiency_2015}.
In conclusion, the average magnetisation is \textit{not} generally enough to account for the field of distant regions when the sample has macroscopic magnetic order. For some macroscopically ordered samples, standard demagnetisation tensors and PBC models are simply inapplicable. 

We also note that for uniform magnetisation rather than \cref{eq:M_pathological}, the $\theta$-integral in \cref{eq:M_pathological_calc} is zero instead of $\sqrt{2}$, thus removing the logarithmic singularity. This suggests that a scale invariant demagnetisation field is specific to uniformly magnetised systems, though we cannot exclude other special magnetic orderings having this property.

\section*{Acknowledgement}
This work was supported by the Carlsberg Foundation Semper
Ardens Advance project CF24-0920 entitled ``Novel magnets through interdisiplinarity and nanocomposites'' and by the Independent Research Fund Denmark, grant ``Magnetic Enhancements through Nanoscale Orientation (METEOR)'', 1032-00251B and grant ``The next generation permanent magnet for the green transition'', 4307-00105B.

\bibliographystyle{unsrt}
\bibliography{references}


\end{document}